\documentclass[%
 reprint,
superscriptaddress,
nobalancelastpage,
 amsmath,amssymb,
 aps,
]{revtex4-2}

\usepackage{graphicx}
\usepackage{dcolumn}
\usepackage{bm}
\usepackage{siunitx}

\begin{document}

\preprint{APS/123-QED}

\title{Fabrication of oriented NV center arrays in diamond via femtosecond laser writing and reorientation}

\author{Kai Klink}
 \affiliation{Department of Physics and Astronomy, Michigan State University, East Lansing, MI 48824, USA}
\author{Andrew Raj Kirkpatrick}%
 \affiliation{Department of Physics and Astronomy, Michigan State University, East Lansing, MI 48824, USA}
\affiliation{Department of Electrical and Computer Engineering, Michigan State University, East Lansing, MI 48824, USA}%
\author{Yukihiro Tadokoro}
 \affiliation{Toyota Research Institute of North America, Ann Arbor, MI 48105, USA}
\author{Jonas Nils Becker$^\ddag$}
    \email{becke183@msu.edu}
    \affiliation{Department of Physics and Astronomy, Michigan State University, East Lansing, MI 48824, USA}
    \affiliation{Coatings and Diamond Technologies Division, Center Midwest (CMW), Fraunhofer USA Inc, East Lansing, MI 48824, USA}
\author{Shannon Singer Nicley$^\ddag$}
    \email{nicleysh@msu.edu}
    \affiliation{Department of Electrical and Computer Engineering, Michigan State University, East Lansing, MI 48824, USA}
    \affiliation{Coatings and Diamond Technologies Division, Center Midwest (CMW), Fraunhofer USA Inc, East Lansing, MI 48824, USA}
    \affiliation{Department of Chemical Engineering and Materials Science, Michigan State University, East Lansing, MI 48824, USA}
    
    \altaffiliation{These authors contributed equally and share last authorship}

\date{\today}

\begin{abstract}
Nitrogen-vacancy (NV) centers in diamond are widely recognized as highly promising solid-state quantum sensors due to their long room temperature coherence times and atomic-scale size, which enable exceptional sensitivity and nanoscale spatial resolution under ambient conditions. Ultrafast laser writing has demonstrated the deterministic spatial control of individual NV$^-$ centers, however, the resulting random orientation of the defect axis limits the magnetic field sensitivity and signal contrast. Here, we present an all-optical approach for reorienting laser-written NV$^-$ centers to lie along a specific crystallographic axis using femtosecond laser annealing. This technique enables the creation of spatially ordered  NV$^-$ arrays with uniform orientation, for enhancing performance for quantum magnetometry. We achieve deterministic alignment along the optical axis in both (100)- and (111)-oriented diamond substrates, paving the way for scalable, high-performance quantum devices based on orientation-controlled NV$^-$ centers. 
\end{abstract}

\maketitle


\section{\label{sec:intro}Introduction}

Among the various platforms for quantum technologies, the negatively charged nitrogen vacancy (NV$^-$) center in diamond is particularly attractive due to their operation at room temperature, solid-state integration, and compatibility with advanced nanofabrication techniques. These features make them uniquely suited for scalable quantum technologies, including  quantum information processing, sensing, and nanoscale imaging. The NV$^-$ center, a point defect consisting of a substitutional nitrogen atom adjacent to a vacant site in a diamond lattice, possesses electron spin states that can be optically initialized, manipulated and read out, even at room temperature \cite{jelezko_observation_2004}. These characteristics enable a wide range of applications,  including magnetometry \citep{nv_magnetometry_2008,katsumi_high_2025}, electrometry \citep{dolde_electric-field_2011}, and thermometry \citep{neumann_high-precision_2013}. They are also highly versatile, and schemes have been reported allowing for their use as sensitive magnetometers in zero bias field \citep{zheng_zero_2019} and without the use of microwave excitation \citep{wickenbrock_microwave_2016, buergler_all_2024}.  Recent demonstrations on the transfer of the spin polarization to proximal nuclear spins to create a spin register have also shown incredible promise for the future of these centers, with spin coherence times exceeding several seconds \cite{bradley_ten_2019,vandeStolpe_mapping_2024}. A key advantage of using NV$^-$ centers lies in their nature as atomic-scale point defects in a solid-state host. Their minute size allows for nanometer scale precision in sensing, for example allowing for the imaging of magnetic nanostructures and currents in microelectronic circuits.

Such applications, however, require deterministic placement of individual NV$^-$ centers with minimal damage to the surrounding lattice. This has been accomplished using ultrafast laser writing, which allows NV$^-$ centers to be fabricated with high spatial precision \citep{chen_laser_2017}. The laser writing process consists of an initial seed pulse that produces vacancies in the lattice via multiphoton ionization, followed by a train of diffusion pulses, which cause vacancies to migrate until an NV$^-$ center is formed. This method has been used to fabricate NV$^-$ centers with an in-plane positioning accuracy of 33 nm \citep{chen_laser_2019}.

\begin{figure*}[t]
\includegraphics[width=1\textwidth]{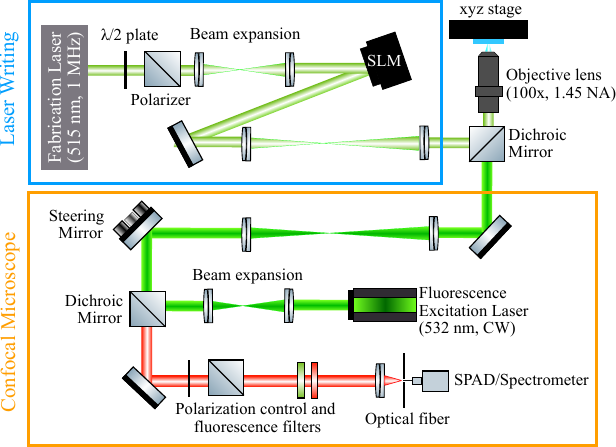}
\caption{\label{fig:optical_diagram}Schematic diagram of the ultrafast laser fabrication system with \textit{in situ} confocal microscope. Ultrafast 515 nm laser pulses (light green) are produced by a Yb:KYW laser and are corrected for aberrations using a spatial light modulator. The beam-scanning confocal fluorescence microscope excites with a 532 nm CW laser (dark green) and collects fluorescence (red) from the NV$^-$ center onto a single photon detector. Both the writing system and the microscope share a common objective lens onto the sample, which is illuminated in transmission for widefield microscopy onto a CMOS camera.}
\end{figure*}

However, laser writing results in NV$^-$ centers oriented along any of the four possible axes, the $[111]$, $[1\bar 1 \bar 1]$, $[\bar 1 \bar 1 1]$, or $[\bar 1 1 \bar 1]$ axes. This is undesirable for magnetometry applications that utilize ensembles of NV$^-$ centers, as each NV$^-$ center is sensitive only to the component of the magnetic field parallel to its orientation axis. This results in an orientation-dependent Zeeman splitting such that only a single orientation can be resonantly addressed at once. Other orientations then contribute to the background, decreasing the contrast of measurements. Preferential orientation of NV$^-$ centers along two axes has been demonstrated to increase the sensitivity of magnetometry with ensembles by a factor of 2 \citep{pham_enhanced_2012}. CVD growth has been used to produce NV$^-$ centers with preferential orientation along two axes in diamonds with (110)- \citep{edmonds_production_2012} and (100)-oriented surfaces \citep{pham_enhanced_2012}, and along a single axis for (111)- \citep{lesik_111_2014, michl_perfect_2014, fukui_perfect_2014}  and (113)-oriented surfaces \citep{lesik_113_2015}. To date, however, there has not been a method demonstrated for the production of well localized \textit{single} NV centers with a specific orientation. The lack of post-fabrication orientation control has been an obstacle to scaling NV-based sensors, particularly in applications requiring uniform spin response such as vector magnetometry and high-contrast imaging of biological systems. 

The NV$^-$ center emits fluorescence (PL) via a linear combination of two orthogonal dipole moments in the plane normal to the NV center orientation axis. The dipoles are energetically degenerate, resulting in a nominally circular polarization pattern. However they can contribute differently to the overall PL signal when analyzed through a polarizer in the laboratory reference frame due to their respective orientations relative to the diamond surface. Hence, polarization patterns can be used to characterize the NV orientation within the diamond lattice \cite{dolan_complete_2014}. Due to the distinct dipole emission patterns in combination with diamond's high refractive index, orientation also affects the luminescence intensity, and the orientation can also be determined by observing this change in intensity \cite{peng_all_2024}.

Here, we combine several of these ideas and introduce a  femtosecond laser annealing technique that allows for reorientation of individual laser written NV$^-$ centers. By selectively dissociating and reforming centers optically, all NV$^-$ orientations are produced stochastically. By combining this with \textit{in situ} orientation detection via fluorescence polarization analysis, this process then enables the selection of a desired crystallographic orientation, allowing for fully deterministic spatial and orientational control of single NV$^-$ centers.

\section{Methods}

NV$^-$ centers are fabricated using a home-built aberration-corrected ultrafast laser writing system with an \textit{in situ} confocal microscope, as shown in Figure \ref{fig:optical_diagram}. The system uses 270 fs pulses at 515 nm generated by a Yb:KYW laser (Light Conversion Pharos PH2). Pulse energy is controlled using a $\lambda/2$ waveplate (Thorlabs WPHSM05-514) and a Glan-laser polarizer (Thorlabs GL5). The beam is then expanded to fill a spatial light modulator (SLM) (dual-band Meadowlark HSP 1920-500-1200), which applies aberration corrections to compensate for the significant spherical aberration caused by diamond's high refractive index \cite{simmonds_three_2011}. The corrected beam is then focused into the diamond using a 1.45 NA oil immersion objective lens (Olympus MPlanApo N 100x). The sample is mounted on high-precision translation stages (Zaber LDM060C for x-y, LDM040C for z; accuracy \SI{1}{\um}, repeatability $<$\SI{80}{\nm}) and widefield transmission imaging is performed using a CMOS camera (Thorlabs CS165MU/M). 

The confocal fluorescence microscope shares the same objective lens as the laser writing system. Non-resonant excitation is provided by a 532 nm continuous-wave laser (Spectra Physics Millennia eV 15) which is scanned across the sample using galvanometric mirrors (Thorlabs QS7XY-AG) enabling spatially independent excitation relative to the fabrication focus. The emitted fluorescence is separated from both the excitation and fabrication wavelengths using a 550 nm longpass dichroic mirror (Thorlabs DMLP567), then coupled into an optical fiber for confocal optical sectioning. The collected fluorescence signal is directed either to a single-photon avalanche detector (SPAD) (Excelitas SPCM-AQRH-14), a spectrometer (Princeton Instruments SpectraPro HRS-750 with Blaze 400HR eXcelon camera) or a pair of SPADs arranged in a Hanbury Brown and Twiss (HBT) interferometer configuration for photon autocorrelation measurements. Both the laser writing system and confocal microscope are controlled via custom software written for a Field Programmable Gate Array (FPGA) (NI USB-7845R) with a user interface developed in LabVIEW and Python. 

NV$^-$ centers fabrication was performed on diamond substrates from two sources: a crystal  with a (111)-oriented surface purchased from Flawless Technical Diamonds (FTD), in which a 3$\times$3 array of NV$^-$ centers was created, and a (100)-oriented diamond substrate provided by Great Lakes Crystal Technologies (GLCT), where a single reoriented NV$^-$ center was produced. In both samples, NV$^-$ centers were fabricated using an initial 270 fs seed pulse (1.47 nJ at 515 nm) to generate vacancies and interstitial carbon atoms via multiphoton ionization. This was followed by a 200 kHz train of 1.19 nJ diffusion pulses which mobilized vacancies until one combined with a substitutional nitrogen atom to form an NV$^-$ center. Fluorescence intensity is monitored during this process, and the pulse train is terminated upon detection of an NV$^-$ signal.  

Polarization analysis of the NV fluorescence for orientation determination was achieved using a $\lambda/2$ waveplate (Thorlabs WPHSM05-694) in front of a polarizing beam splitter (Thorlabs PBS252). By rotating the half-wave plate, the full polarization profile of the emitted light was measured. The different projections of the NV center's emission dipoles onto the optical pupil enabled the identification of orientation based on the polarization pattern.  All four orientations can be differentiated in this manner for diamond with a (111)-oriented surface, whereas only two sets of orientations are distinguishable for diamond with a (100)-oriented surface. If the measured orientation does not match the desired orientation, an additional annealing pulse train was applied causing reorientation of the NV$^-$ center. Polarization measurements are repeated to identify the new orientation. This process is repeated until the desired orientation is observed. The fabricated NV$^-$ centers are then characterized using photoluminescence excitation spectroscopy and photon autocorrelation measurements. Occasionally, when reorientation does not readily occur after applying diffusion pulses for several minutes, a second seed pulse is applied to help the process.

\section{Results and Discussion}
\begin{figure}[t]
\includegraphics[width=1\columnwidth]{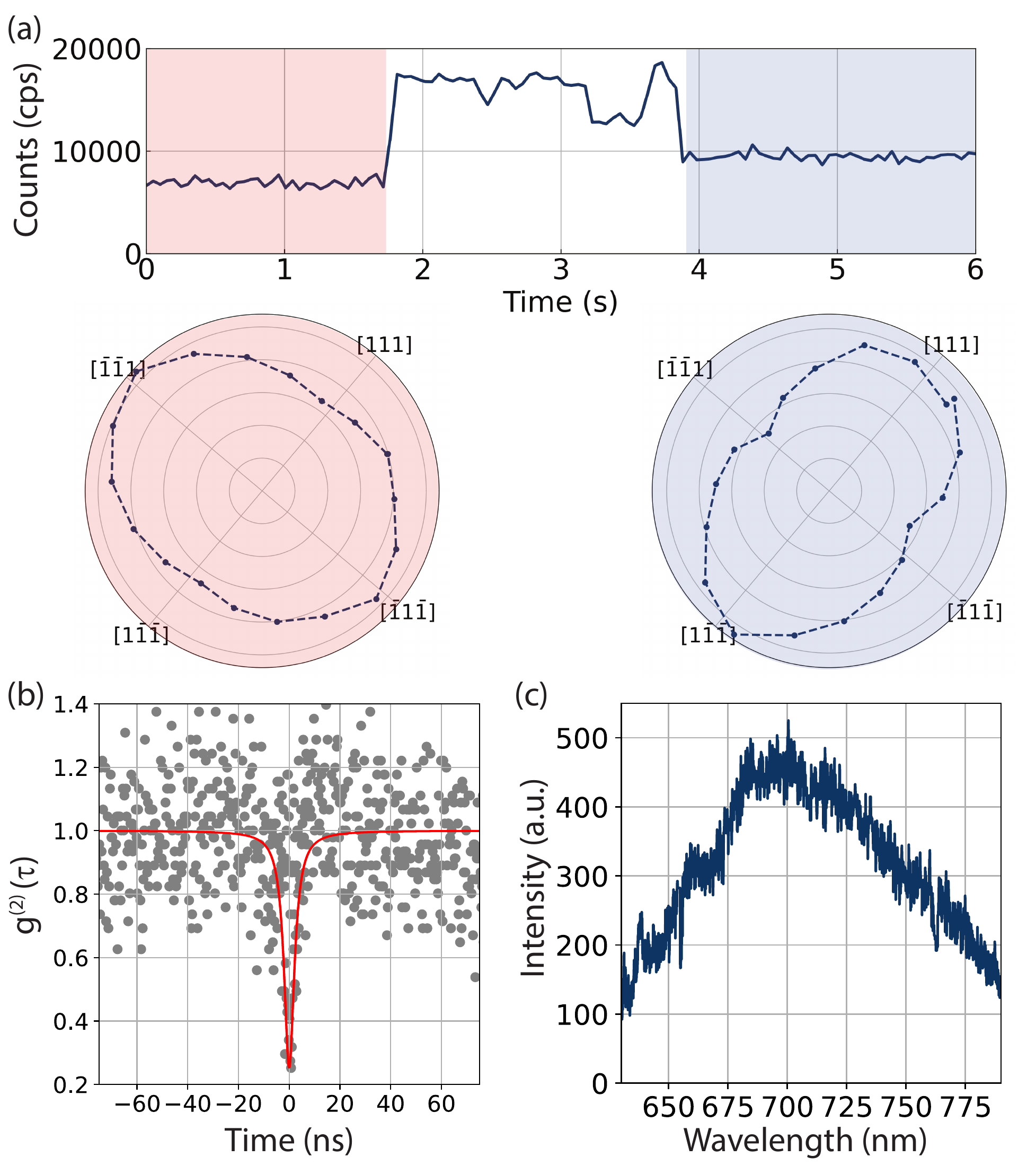}
\caption{\label{fig:glct} \textbf{(a)} Fluorescence trace of an NV$^-$ center during reorientation in (100)-orientated diamond. Initially an NV$^-$ center is created along either of the, polarization degenerate, $[111]$ or $[1\bar{1}\bar{1}]$ directions, as shown by the red-shaded polarization map. The diffusion pulse train is active between 1.8 and 3.9 s on the fluorescence trace. After some fluctuation in the measured fluorescence during diffusion, the NV$^-$ orientation is probed resulting in the blue-shaded polarization map. This demonstrates a NV$^-$ orientation change between the polarization degenerate orientations of $[\bar{1}\bar{1}1]/[\bar{1}1\bar{1}]\rightarrow[111]/[1\bar{1}\bar{1}]$. The fluorescence is confirmed to be from a single NV$^-$ by \textbf{(b)} Second-order photon autocorrelation measurement and \textbf{(c)} Room temperature fluorescence spectrum.}
\end{figure}

First, a single NV$^-$ center was fabricated and reoriented in a (100)-oriented diamond substrate. The initial polarization signature of the center is shown in red in Fig. \ref{fig:glct}(a). Prior to reorientation, the half-waveplate in the fluorescence collection path was rotated to minimize the detected fluorescence signal. This provides polarization-dependent fluorescence contrast between the two distinguishable orientation classes observable in (100)-oriented diamond, allowing for quick reorientation detection. Fig. \ref{fig:glct}(a) shows the time-resolved fluorescence signal during the reorientation process. The diffusion pulse train was activated at 1.8 s. A transient dip in fluorescence was observed just after the 3 s mark, followed by recovery to a higher fluorescence level. The diffusion pulse train was then terminated. The increased fluorescence observed after the reorientation process is consistent with the change in polarization and thus increased transmission through the polarization analyzer. The new orientation was confirmed by a full measurement of the polarization pattern as shown shaded in blue in Fig. \ref{fig:glct}(a). Single photon emission was verified by photon autocorrelation, yielding a background-corrected $g^{(2)}(0)=0.25$, as shown in Fig. \ref{fig:glct}(b). A characteristic NV$^-$ fluorescence spectrum was measured on the reoriented emitter as shown in Fig. \ref{fig:glct}(c). In (100)-oriented diamond, NV$^-$ centers fall into two optically distinguishable orientation classes. This partial control over orientation enables the creation of NV  arrays with only two orientations, potentially doubling the sensitivity in magnetometry applications \cite{pham_enhanced_2012}.

\begin{figure*}[t]
\includegraphics[width=1\textwidth]{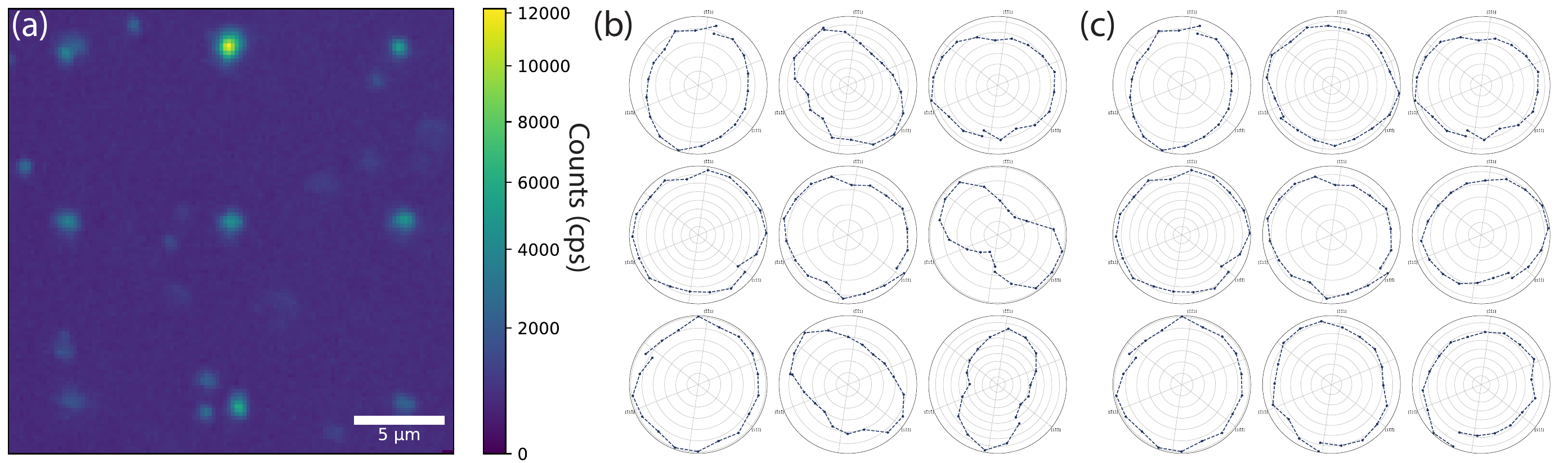}
\caption{\label{fig:ftd}\textbf{(a)} Confocal fluorescence image of laser written array of 9 NV$^-$ centers aligned parallel to the optical axis in (111)-oriented diamond. \textbf{(b)} Initial polarization maps of the NV$^-$ centers before reorientation, demonstrating random orientation. \textbf{(c)} Polarization maps for each NV$^-$ center in the array after reorientation to the optical axis, which identify all the NV$^-$ centers as being aligned along the optical axis in (111)-oriented diamond.}
\end{figure*}

To fully exploit the abiliy to distinguish all four NV$^-$  orientations, we extended our method to a (111)-oriented diamond substrate. An array of nine NV$^-$ centers was fabricated with \SI{10}{\micro\meter} pitch at a depth of \SI{20}{\micro\meter}, as shown in Fig. \ref{fig:ftd}(a). Polarization analysis (Fig. \ref{fig:ftd}(b)) revealed that five of the nine fabricated NV$^-$ centers were initially oriented along the (111) crystallographic direction (i.e. parallel to the optical axis), while the remaining four NV$^-$ centers were distributed among two of the three other possible orientations. The fourth orientation of the centers was absent, due to the limited number of centers in the fabricated array (but has been observed in other fabrication runs). 

The reorientation process itself was identical to the (100) sample, with the exception that the polarization was not biased to minimize counts. Instead a full polarization map was taken when fluctuations in the fluorescence level occurred. This was because the difference in fluorescence intensity using a single linear polarization analyzer between the four orientations does not provide a clear contrast in (111)-oriented diamond. Each NV$^-$ center was reoriented to the (111) direction parallel to the optical axis (vertical), which was confirmed by polarization measurements shown in Fig. \ref{fig:ftd}(c). Note that the speed and simplicity of this process could in principle be improved further by splitting the fluorescence into multiple paths with separate polarization analyzer set to the individual orientation axes of (111)-oriented material. Comparison of the PL intensity in the individual arms can then provide quick feedback on the NV orientation without the need to measure a full polarization pattern.

Such an NV$^-$ array aligned along the optical axis has two distinct advantages over an unaligned array. Firstly, it allows for each member of the array to be coherently addressed simultaneously under the same magnetic field. This would in principle lead to a factor of 4 increase in sensitivity when compared to a randomly oriented array. Secondly, NV$^-$ centers aligned along the optical axis emit more light into a smaller numerical aperture of an objective lens and less light is lost during total internal reflection, ensuring both maximal light collection and reduction in non-symmetric aberrations accumulated at the diamond surface, improving sensitivity due to its enhanced photon collection. An array pitch of \SI{10}{\micro\meter} was chosen due to evidence in the literature that the diffusion process can influence an area of diamond much larger than the focal spot \cite{cheng_laser_2024}. Further investigations of the diffusion process are needed to minimize potential unwanted effects on neighboring NVs and reduce emitter spacing.

Two potential, non-competing, mechanisms explain how reorientation occurs. Firstly, the initial NV$^-$ center is dissociated during annealing, allowing a vacancy to migrate and reform an NV$^-$ center in a new orientation \cite{oberg}. This occurrence is proposed within the literature as the dominant mechanism by which NV centers migrate in diamond under thermal annealing, however it is yet to be determined if ultrafast laser diffusion is able to exceed the energy barrier required to dissociate an NV$^-$ center in this process. Although this mechanism seems promising considering the reduction in fluorescence during diffusion demonstrated in Fig. \ref{fig:glct}(a), there may be other mechanisms that can cause an NV$^-$ center to stop fluorescing without dissociation. Hybridization between the NV$^-$ center's e-manifold and the $/pi$-bonds of a (100)-split carbon self-interstitial has been demonstrated and could reduce the fluorescence emission from the NV$^-$ center \cite{kirkpatrick_ab_2024}. Therefore, it is possible that the reorientation process is substantially different from the thermal mechanism, with the initial NV$^-$ center continuing to exist initially but as a dark modified defect complex involving additional entities such as interstitials or vacancies, followed by a dissociation step of this intermediate defect resulting in a reoriented NV center. Further evidence for such a process is the occasional need to reseed after prolonged, unsuccessful reorientation attempts. This is likely due to a lack of species in the NV vicinity which catalyze the reorientation process, most likely additional vacancies.  

In conclusion, we have demonstrated a fully optical method for the deterministic orientation of single NV$^-$ centers with high spatial resolution in diamond using femtosecond laser annealing. Using this method, we demonstrated the feasibility of fabricating  oriented arrays of NV$^-$ centers in diamonds in both (100)- and (111)-oriented diamond substrates. Being an all-optical process, ultrafast laser writing provides unique control of the defect formation process and defect properties. Such techniques provide invaluable flexibility in fabricating optimized spin-photonic systems. Future work in this field may extend these methods to other defects in diamond or other materials system, potentially in combination with the preparation of suitable precursor materials. Moreover, further optimization is needed to make this process suitable for writing of shallow defects and direct writing into nanostructured diamond. Finally, a better understanding of the microscopic processes underpinning defect creation and dynamics in femtosecond laser fabrication via the development of theoretical models in combination with ultrafast spectroscopy is needed for further optimizations.

\section*{Conflict of Interest Statement}
The authors declare that the research was conducted in the absence of any commercial or financial relationships that could be construed as a potential conflict of interest.

\section*{Author Contributions}
KK: Writing – original draft, Software, Investigation, Visualization; ARK: Writing – original draft, Visualization, Software, Investigation, Conceptualization; YT - Writing – review \& editing, Resources, Project administration, Funding acquisition, Conceptualization; JNB: Writing – review \& editing, Supervision, Project administration, Funding acquisition, Conceptualization; SSN - Writing – review \& editing, Supervision, Project administration, Funding acquisition, Conceptualization.

\section*{Funding}
 This work was supported through a collaborative research agreement funded through Toyota Motor Engineering \& Manufacturing, North America, Inc.

\section*{Acknowledgments}
J.N.B is supported by the Cowen Family Endowment. K.K. acknowledges support from a Lawrence W. Hantel Endowed Fellowship at MSU. We thank Great Lakes Crystal Technologies (GLCT) for providing a suitable diamond sample for this work and Paul Quayle for helpful comments.

\section*{Data Availability Statement}
The original contributions presented in the study are included in the article, further inquiries can be directed to the corresponding authors.

\bibliography{main}

\end{document}